\documentclass[footinbib,superbib,eqsecnum,prb,nobibnotes]{revtex4}

\usepackage{hyperref}
\usepackage{amsmath,amsthm,amssymb}
\usepackage[latin1]{inputenc}
\usepackage{epic,curves}

\newtheorem{defn}{Definition}
\newtheorem{lemm}[defn]{Lemma}
\newtheorem{prop}[defn]{Proposition}
\newtheorem{thm}[defn]{Theorem}

\def \opmh {\omega^{\pm 1/2}}
\def \omph {\omega^{\mp 1/2}}

\newcommand{\fpm}{ f ^\pm }

\newcommand{\mkraum}{ \rbb^{s+1} }

\DeclareMathOperator{\rank}{rank}
\DeclareMathOperator{\img}{img}
\DeclareMathOperator{\lspan}{span}
\DeclareMathOperator{\supp}{supp}

\def \pv {\mathbf{p}}
\def \pvq {\mathbf{p}^2}
\def \xv {\mathbf{x}}

\newcommand{ \cinfty } {  \ccal^\infty (\hcal) }
\newcommand{ \cinftys } {  \ccal^\infty(\Sigma) }
\newcommand{\lnorm}[2]{  \| #1 \|^{( #2 )}  }
\def \PhiFH { \Phi_{ \mathrm{FH} }  }
\def \lcs { \Psi }
\newcommand{\lcssample}{ \psi }
\newcommand{\psmap}{\Xi}

\newcommand{\wickprod}[1] {  : \!\! #1 \!\! : }

\newcommand{\idop}{\openone}

\def \cbb{\mathbb{C}}
\def \nbb{\mathbb{N}}
\def \rbb{\mathbb{R}}

\def \ccal {\mathcal{C}}

\def \ecal {\mathcal{E}}
\def \fcal {\mathcal{F}}

\def \hcal {\mathcal{H}}

\def \kcal {\mathcal{K}}
\def \lcal {\mathcal{L}}

\def \ocal {\mathcal{O}}
\def \pcal {\mathcal{P}}

\def \scal {\mathcal{S}}

\def \vcal {\mathcal{V}}

\def \afk  {\mathfrak{A}}
\def \bfk  {\mathfrak{B}}

\def \ffk  {\mathfrak{F}}

\def \cdotarg { \, \cdot \, }

\def \< {\langle}
\def \> {\rangle}

\def \half {\frac{1}{2}}

\def \equivalent {\quad \Leftrightarrow \quad}

\def \st {^\ast}
\def \restrict {\lceil}

\newcommand{\etskp}[2]{\langle #1 | #2 \rangle }
\newcommand{\hrskp}[2]{( #1 | #2 ) }
\newcommand{\bighrskp}[2]{\big( #1 \,\big|\, #2 \big) }

\def \boundedops {\bfk(\hcal)}

\def \dim {\mathrm{dim}\,}

\def \tracenorm#1 { \| #1 \| _1 }
\def \hsnorm#1 { \| #1 \| _2 }

\numberwithin{defn}{section}
\renewcommand{\thedefn}{\arabic{section}.\arabic{defn}}

\begin{document}

\title{Phase space properties and 
the short distance structure in quantum field theory}
\author{Henning Bostelmann}
 \email[Electronic mail: ]{bostelm@theorie.physik.uni-goettingen.de}
\affiliation{%
Universität Göttingen, Institut für Theoretische Physik, 
37077 Göttingen, Germany
}
\date{April 3, 2005}
%

\begin{abstract}

The paper investigates relations between the phase space structure of a quantum
field theory (``nuclearity'') and the concept of pointlike localized fields.
Given a net of local observable algebras, a phase space condition is introduced
that allows a very detailed description of the theory's
field content. An appendix discusses noninteracting models as examples.
\end{abstract}

\maketitle

\hyphenation{pro-ducts}

\section{Introduction}

Quantum fields are a basic ingredient of relativistic quantum physics.
It is common to express almost all aspects of a theory,
including the dynamics, in terms
of these pointlike localized fields and related concepts
(Lagrangians, field equations, path integrals).
While this {\em ansatz} has proved to be very fruitful for the construction of models
and in perturbation theory, it contains technical pitfalls which make
its consistent handling rather difficult.

The main problem arises from Heisenberg's uncertainty relation,
which says that measurements with sharp localization in configuration space
are completely delocalized in momentum space, i.e., they show
a singular high energy behavior.
This is reflected in peculiar mathematical properties of quantum fields:
They cannot be defined as operators, their products do not exist,
and nonlinear field equations require subtle limiting processes which
are difficult to control.

Thus, the concept of pointlike localized fields may be regarded as an
over-idealization. Since it is well known that physical operations
in finite space-time regions (rather than at points)
do not show the
singular behavior mentioned above, it seems worth while to study how
pointlike fields emerge from such a physically more meaningful setting,
and in that way to gain more information about the fields' properties.

A framework for such an investigation
is provided by algebraic field theory,\cite{Haa:LQP}
where operations localized in an open space-time region $\ocal$ form an algebra
$\afk(\ocal)$ of bounded operators.
It is possible to encode all physically relevant
properties (locality, covariance, positivity of energy) in terms of these algebras
and their transformation behavior under the Poincaré group, without reference
to pointlike structures.

While it has extensively been discussed in the literature how to pass
to the local algebras from a given field-theoretic model,\cite{BorYngQuotes}
the reverse step is less well understood. Heuristically,
pointlike fields $\phi(x)$  should be derived as limits of operations
localized in regions shrinking to the point $x$.\cite{FreHer:pointlike_fields,FreJor:products}
However, the details of this
limit depend on the high energy behavior of the fields.
Following Fredenhagen and Hertel, \cite{FreHer:pointlike_fields}
we will focus on fields satisfying {\em polynomial $H$-bounds}, i.e., fulfilling
for some $\ell>0$,
\begin{equation}
 \|R^{\ell} \phi(x) R^{\ell} \|  < \infty ,\quad
 \text{where } R = (1+H)^{-1};
\end{equation}
$H$ denotes the (positive) Hamiltonian.
These $H$-bounds are satisfied in all models constructed so far,
more generally in theories which satisfy a sharpened version of the
Osterwalder-Schrader axioms.\cite{DriFro:reconstruction}
It has been shown in Ref.~\onlinecite{FreHer:pointlike_fields} 
that fields of this kind can be recovered
from the local algebras by means of the formula
\begin{equation}
  R^{\ell} \phi(x) R^{\ell} \in \bigcap_{\ocal \ni x} \overline{ R^{\ell} \afk(\ocal) R^{\ell} }^w ,
\end{equation}
where $\overline{\quad}^w $ denotes the weak closure.\cite{generalized_bounds}
So one can calculate the field content of an algebraic theory; however,
many details about the fields' properties remained unclear, e.g., the question
whether the fields possess finite spin, or the existence of product expansions.

To gain more insight in this respect, we will make use of an additional
property of quantum field theories, known as
{\em phase space conditions}.
These reflect the heuristic expectation that the number of local degrees of
freedom of a physical system is limited: A system restricted both in momentum
and configuration space, thus being localized in a finite phase space volume,
will possess only a finite number of independent physical states. (This is easily understood
by appealing to a semiclassical picture, such as Bohr-Sommerfeld quantization.)
Clearly, the notion of a ``finite phase space volume'' must be handled with care;
however, such properties can be formulated in a mathematically precise way
in terms of {\em compactness} or {\em nuclearity} 
conditions,\cite{HaaSwi:compactness,BucWic:causal_independence,BucPor:phase_space}
which lead to interesting structural results on the system's physical
behavior.\cite{CompactnessResults}
For the sake of concreteness, consider the following compactness condition:\cite{FreHerCompactUnpub}
Let $\hcal$ be the theory's underlying Hilbert space,
$\Sigma$ the set of weak-$\ast$-continuous functionals on $\boundedops$
(``state space''), $P_H(E)$ the spectral projectors of $H$, and $\ocal_r$
the standard double cone of radius $r$ centered at the origin. We require that
for any $E,r>0$, the map
\begin{align} \label{psmapFix}
  \psmap_{E,r} : \quad & P_H(E) \Sigma P_H(E)  \to \afk(\ocal_r)_\ast\,, \\
  & \sigma \mapsto \sigma \restrict \afk(\ocal_r)     \notag
\end{align}
is compact, i.e., can be expanded into a series of rank-1 operators
in the norm topology.
This condition was verified explicitly in noninteracting models.\cite{BucPor:phase_space}

The connection of these properties with the point field structure was
first realized by Haag and Ojima:\cite{Haa:models,HaaOji:germs}
Following the compactness condition mentioned above, the image of the map $\Xi_{E,r}$
is ``almost finite-dimensional'' (up to some degree of precision).
If we assume that this finite-dimensional space does not change
in the limit $r \to 0$ (which may be expected in theories that are
dilation invariant in their short distance limit), then we have found
finitely many independent states that represent the short
distance structure of the theory, and one may hope to interpret a corresponding dual basis
as pointlike localized quantum fields located at $x=0$.

The present paper aims at putting these heuristic ideas into a precise mathematical form.
We introduce a phase space condition, specifically sensitive in the short
distance limit, that selects a relevant class of models with ``regular'' short
distance behavior. Its implications can be summarized as follows.

\begin{enumerate}

\renewcommand{\theenumi}{(\arabic{enumi})}
\renewcommand{\labelenumi}{\theenumi}

\item
The field content of these theories can be determined, with detailed
results on the singular properties of the fields and their
connection to the theory in finite regions.
All fields can be shown to comply with the Wightman axioms; moreover,
they transform covariantly both under unbroken and spontaneously broken
symmetries.

\item
Operator product
expansions \cite{OPE} can be established rigorously
with specific convergence properties,
and allow for a model-independent,
nonperturbative definition of normal products in the sense of
Zimmermann. \cite{Zim:field_equations_phi4,Zim:Brandeis} The field content
includes all composite fields.

\end{enumerate}

In this paper, we shall restrict ourselves to the first named aspect; the latter
point will be treated in another publication.\cite{Bos:products-article}
Section~\ref{phspsec} introduces the phase space condition which lies at the root
of our analysis. Using this condition, we construct the field content of
the algebraic theory in Sec.~\ref{fieldcontentsec}, while symmetry aspects
and the proof of Wightman's axioms are discussed in Sec.~\ref{symmsec}. We end
with a brief outlook on generalizations in Sec.~\ref{outlooksec}. In the Appendix,
we show that the phase space condition is fulfilled in models from free field theory.

The present paper is based on the author's thesis.\cite{Bos:Operatorprodukte}
It presents an abbreviated and somewhat simplified version of material
developed there; for further details and additional technical and mathematical aspects,
the reader is referred to the original work.

\section{A phase space condition} \label{phspsec}

This section will introduce the specific phase space condition
which characterizes the short distance structure of a quantum field theory.
First of all, we shall briefly define our mathematical setting.

We use the framework of Local Quantum Physics.\cite{Haa:LQP}
Our field-theoretic
model is given by means of a net $\ocal \mapsto \afk(\ocal)$, which associates
with any open space-time region $\ocal \subset \mkraum$ a von Neumann algebra
$\afk(\ocal) \subset \boundedops$, where $\hcal$ is some fixed Hilbert space.
We assume isotony and locality.
In most cases, it will suffice to consider the algebras
$\afk(r) := \afk(\ocal_r)$ associated with standard double cones $\ocal_r$
of radius $r$ centered at the origin.
We will restrict our attention to
the vacuum sector; i.e., we assume a strongly continuous unitary representation
$U(x,\Lambda)$ of the connected Poincaré group on $\hcal$
that has a geometric action on the net $\afk$ {\em (covariance)},
and a vector $\Omega \in \hcal$ (the {\em vacuum}), unique up to a phase,
which is invariant under all $U(x,\Lambda)$.
For the translation part of this representation, $U(x) = \exp(i P_\mu x^\mu)$,
we will assume the {\em spectrum condition,} i.e., the joint spectrum of
its generators $P_\mu$ falls into the closed forward light cone $\overline{\vcal}_+$;
here $H = P_0 \geq 0$ is the Hamiltonian. 
We also require the net $\afk$ to act {\em irreducibly} on $\hcal$, in the sense that
$\cup_\ocal \afk(\ocal) \Omega$ is dense in $\hcal$, where the union runs over 
bounded regions $\ocal$ only.
$\Sigma := \boundedops_\ast$
is the set of weak-$\ast$-continuous
functionals on $\boundedops$, the positive normed elements of which represent
the physical states. Both $\boundedops$ and $\Sigma$ will be considered with
their norm topology unless otherwise noted.

In analogy to Ref.~\onlinecite{FreHer:pointlike_fields},
we will often use the space of polynomially energy-damped functionals,
\begin{equation}
 \cinftys := \bigcap_{\ell>0} R^{\ell} \Sigma R^{\ell} \subset \Sigma,
\end{equation}
where $R = (1+H)^{-1}$ as before.
On this space, we define the norms
$\lnorm{\sigma}{\ell} := \| R^{-\ell} \sigma R^{-\ell}\|$
$(\ell>0)$,
and equip $\cinftys$ with the locally convex (metrizable) topology
of simultaneous convergence with respect to all these
norms.  As is easily seen, a linear map $\varphi$ from $\cinftys$ to
some Banach space ($\ecal$,$\|\cdotarg\|_\ecal$) is
continuous with respect to this topology if and only if
\begin{equation}
  \lnorm{ \varphi }{\ell}
  := \sup_{\sigma} \frac{ \| \varphi(\sigma) \|_\ecal }{ \lnorm{ \sigma }{ \ell } }
  < \infty \quad \text{for some } \ell>0.
\end{equation}
We set $\lcs := \lcal( \cinftys, \Sigma)$, the space of continuous
linear maps between these spaces. A prominent element of $\lcs$ is
the inclusion
\begin{align} \label{psmapGlobal}
  \psmap : \quad & \cinftys  \to \Sigma, \\
  & \sigma \mapsto \sigma,     \notag
\end{align}
which bears some analogy with \eqref{psmapFix}. A map $\lcssample \in \lcs$
of finite rank has the form
\begin{equation} \label{finite_rank_sample}
  \lcssample = \sum_{j=1}^{J} \sigma_j \phi_j 
  \qquad
  \text{with } \sigma_j \in \Sigma
  \quad \text{ and } \phi_j \in \cinftys^\ast.
\end{equation}
(This means that the $\phi_j$ are linear forms with $\lnorm{\phi_j}{\ell}
= \|R^{\ell} \phi_j R^{\ell} \| < \infty$ for large~$\ell$.)

For our purposes, it will be critical to control the behavior of the
objects mentioned above in the short distance limit. To achieve this, we
introduce some additional structure. For $\sigma \in \Sigma$, we define
\begin{equation}
  \| \sigma \|_r := \| \sigma \restrict \afk(r) \|
  \quad (r>0).
\end{equation}
Likewise, for $\lcssample \in \lcs$ we consider
\begin{equation}
  \lnorm{\lcssample}{\ell} _r := \lnorm{ \lcssample(\cdotarg) \restrict \afk(r) }{\ell}
  = \sup_{\sigma \in \Sigma} \sup_{A \in \afk(r)}
    \frac{ |\lcssample(R^{\ell} \sigma R^{\ell}) \,(A) | }{ \|\sigma\| \, \|A\| },
\end{equation}
which is finite for sufficiently large $\ell$.
We devote special attention to the question how fast such a seminorm
$\lnorm{ \cdotarg }{\ell} _r $ vanishes in the limit $r \to 0$, and therefore
define for $\gamma \geq 0$,
\begin{equation} \label{deltadef}
  \delta_\gamma (\lcssample) :=
  \begin{cases}
    0 & \text{if } \; r^{-\gamma} \lnorm{\lcssample}{\ell} _r \xrightarrow[r \to 0]{} 0
    \text{ for some $\ell>0$},  \\
    1 & \text{otherwise.}
  \end{cases}
\end{equation}
Each $\delta_\gamma$ induces a pseudometric on $\lcs$.
Equipping $\lcs$ with the pseudometric
\begin{equation}\label{hatdeltadef}
  \hat \delta (\lcssample,\lcssample') := \int_0^\infty e^{-\gamma} \,\delta_\gamma(\lcssample-\lcssample') \,d\gamma \,,
\end{equation}
we have found a topology
on $\lcs$ which describes short-distance convergence
``to all polynomial orders.''

[Note that $\lcs$ is a presheaf over $\rbb^{s+1}$
with respect to $\lcs(\ocal) := \lcs \restrict \afk(\ocal)$, and that the $\delta_\gamma$
are not so much functions on $\lcs$, but rather on the stalk of $\lcs$ at $x=0$;
so we have defined a topology on the stalk.
However, we shall not consider these sheaf-theoretic aspects any further.]

We can now turn to formulating the approximation of $\Xi$ with
operators of finite rank. We start with a net $\ocal \mapsto \afk(\ocal)$
that meets the standard requirements listed at the beginning of this section.
We can analyze the short distance behavior of the theory
with the help of the concepts introduced above by considering the numbers
\begin{equation} \label{ngammadef}
  N_\gamma := \min \big\{ n \; | \; \exists \, \lcssample \in \lcs: \,
                   \rank \lcssample = n, \;
                     \delta_\gamma(\psmap-\lcssample) =0  \big\},
\end{equation}
setting $N_\gamma = \infty$ if the set on the right-hand side is empty.
It is obvious that $N_\gamma \geq N_{\gamma'}$ for $\gamma \geq \gamma'$.
Thus, our theory falls into one out of the following three classes:

\begin{enumerate}

\renewcommand{\theenumi}{(\arabic{enumi})}
\renewcommand{\labelenumi}{\theenumi}

\item    \label{regularcase}
$N_\gamma \to \infty$ for $\gamma \to \infty$,
but $N_\gamma < \infty$ for all $\gamma \geq 0$.
We shall show in the subsequent sections that
the theory then has a nontrivial field content of Wightman fields.
As will be discussed in the Appendix, a large variety of free field theories falls
into this class; one may expect that physically relevant interacting 
models exhibit the same behavior.

\item     \label{trivialcase}
$N_\gamma \leq N$ for some fixed $N < \infty$ and all $\gamma \geq 0$.
In this case, only very few observables ``survive'' in the short distance limit.
In fact, we shall see that in this case necessarily $N_\gamma = 1$ for all $\gamma$,
and that the theory's field content is trivial.
An example for this behavior was discussed by Lutz; \cite{Lut:UVFix}
cf. also the Appendix. More generally, one expects
that models of this kind are {\em not} generated by pointlike fields, but
possibly by other, nonpointlike objects, such as Mandelstam strings or
Wilson loops.

\item     \label{irregularcase}

$N_\gamma = \infty$ for some finite $\gamma \geq 0$. Such theories
have a peculiar complex short distance behavior which does not allow
$\psmap$ to be approximated by maps of finite rank in the specified sense.
Examples are free theories with infinitely many species of particles.
One also expects that theories in $1+1$ dimensions, where the fields
as well as their Wick products have a ``scaling dimension'' of $0$,
are members of this class.

\end{enumerate}

In what follows, we will disregard the latter case and assume that
$N_\gamma$ is finite for all $\gamma$.
We may encode this into a phase space criterion which selects theories of
a sufficiently regular short distance behavior.

\begin{defn}
  A net $\ocal \mapsto \afk(\ocal)$ is said to satisfy the
  {\em microscopic phase space condition} if for every $\gamma \geq 0$,
  there exists a map $\lcssample \in \lcs$ of finite rank such that
  \begin{equation*}
    \delta_\gamma(\psmap-\lcssample) = 0,
  \end{equation*}
  or, equivalently,
  \begin{equation*}
    r^{-\gamma} \lnorm{ \psmap-\lcssample}{\ell} _r \to 0     \quad
    \text{for sufficiently large }\ell>0.
  \end{equation*}
\end{defn}

This condition shows a formal analogy with the well-known compactness
and nuclearity conditions in finite regions:\cite{BucWic:causal_independence,BucPor:phase_space}
It demands that the map $\Xi$ can be 
expanded into a series of rank-1 operators, where the series is meant
to converge in the pseudometric $\hat \delta$ introduced in Eq.~\eqref{hatdeltadef}.

Our task will be to analyze the consequences of the microscopic phase space condition,
especially regarding the theory's point field structure. We will show that
the size of the field content is determined by the ``approximation numbers''
$N_\gamma$, and prove that these fields fulfill certain regularity properties.

\section{Determining the field content} \label{fieldcontentsec}

We will now set out from a net $\ocal \mapsto \afk(\ocal)$
that satisfies the microscopic phase space condition. Our task is
to analyze the point field content of this net. Our line of
analysis can be sketched as follows.

Given $\gamma \geq 0$, let $\lcssample \in \lcs$ be a map of finite rank
that approximates $\psmap$ in the short distance limit: $\delta_\gamma(\Xi-\lcssample)=0$.
Provided that $\rank \lcssample$ is minimal with this property (i.e. $\lcssample$
does not contain redundant terms that do not contribute to the approximation),
one will expect that the image of the
dual map $\lcssample^\ast: \boundedops \to \cinftys^\ast$ consists of
pointlike fields. [Note that $\img \lcssample^\ast$ is spanned by the linear forms $\phi_j$ in
Eq.~\eqref{finite_rank_sample}.]
In fact, we will show that $\Phi_\gamma := \img \lcssample^\ast$ depends on
$\gamma$ only, and that its elements are indeed localized fields.
Furthermore, we will prove that the union of the spaces $\Phi_\gamma$
exhausts the field content as defined by Fredenhagen and Hertel.\cite{FreHer:pointlike_fields}

Let $\gamma \geq 0$ be fixed in the following. We begin our analysis of
the map $\lcssample$ with the following lemma.

\begin{lemm} \label{sigma_not_vanish_lemma}
Let $\lcssample \in \lcs$ such that $\rank \lcssample = N_\gamma$ and
$\delta_\gamma(\psmap-\lcssample) = 0$. Then it holds for all
$\sigma \in \img \lcssample \backslash \{0\}$ that
\begin{equation*}
  r^{-\gamma} \| \sigma \|_r \not \to 0 \qquad
  (r \to 0).
\end{equation*}
\end{lemm}

\begin{proof}
  Assume that the proposition is violated for some $\sigma \in \img \lcssample \backslash \{0\}$.
  We choose a decomposition $\psi = \psi' \oplus  \sigma \phi$,
  where $\phi \in \cinftys^\ast$, $\lcssample' \in \lcs$, and $\rank \lcssample' = N_\gamma-1$.
  Since $r^{-\gamma} \| \sigma \|_r \to 0$ and $\lnorm{\phi}{\ell} < \infty$ for large $\ell$,
  we obviously have $\delta_\gamma( \sigma \phi ) = 0$ 
  and thus $\delta_\gamma(\psmap-\lcssample') =0$, in contradiction
  to the minimality property of $N_\gamma$ [cf. Eq.~\eqref{ngammadef}].
\end{proof}

Certainly, $\lcssample$ is not fixed uniquely by the conditions of Lemma~\ref{sigma_not_vanish_lemma}.
However, we shall show that these conditions fix $\img \lcssample^\ast$. 
To this end, we prove the following.

\begin{lemm}  \label{img_charact_lemma}
  Let $\lcssample$ be as in Lemma~\ref{sigma_not_vanish_lemma}.
  For arbitrary $\sigma \in \cinftys$, the following equivalences hold:
  \begin{equation*}
    \sigma \restrict \img \lcssample^\ast = 0
    \equivalent
    \lcssample(\sigma) = 0
    \equivalent
    r^{-\gamma} \|\sigma\|_r \to 0.
  \end{equation*}
\end{lemm}

\begin{proof}
The first equivalence is obvious. Moreover, 
for $\sigma \neq 0$, the approximation property $\delta_\gamma(\Xi-\psi)=0$
gives $r^{-\gamma} \| (\psmap-\lcssample)(\sigma)\| _r = r^{-\gamma}
\| \sigma - \lcssample(\sigma) \|_r \to 0$. So
$\lcssample(\sigma) = 0 $ implies $ r^{-\gamma} \|\sigma\|_r \to 0$. Conversely,
if $r^{-\gamma} \|\sigma\|_r \to 0$, we conclude  $r^{-\gamma} \|\lcssample(\sigma)\|_r \to 0$,
and thus $\lcssample(\sigma) =0$ according to Lemma~\ref{sigma_not_vanish_lemma}.
\end{proof}

Since the right-hand side of the equivalence in Lemma~\ref{img_charact_lemma}
does not refer to $\lcssample$, the space $\img \lcssample^\ast$ is independent of
$\lcssample$ as well. Thus, we have established the following.

\begin{thm} \label{phi_gamma_def_thm}
Let the net $\afk$ fulfill the microscopic phase space condition.
For each $\gamma \geq 0$, there is a unique subspace $\Phi_\gamma \subset \cinftys^\ast$
of dimension $N_\gamma$ with the following property:
If $\lcssample \in \lcs$, $\rank \lcssample =N_\gamma$, and $\delta_\gamma(\psmap-\lcssample)=0$,
then $\img \lcssample^\ast = \Phi_\gamma$.
\end{thm}

Note that the microscopic phase space condition guarantees such 
a $\lcssample$ to exist for every $\gamma$.
The theorem requires the rank of $\lcssample$ to be ``minimized'' 
($\rank \lcssample = N_\gamma$).
If $\rank \lcssample > N_\gamma$, we can still find a representation
\begin{align} \label{psisplit}
 &\lcssample = \lcssample_{\mathrm{I}} \oplus \lcssample_{\mathrm{II}}\,, \quad \text{where} \notag
 \\
 &\img \lcssample_{\mathrm{I}}^\ast = \Phi_\gamma, \;\quad
 r^{-\gamma} \|\sigma\|_r \to 0 \; \;\forall\, \sigma \in \img\lcssample_{\mathrm{II}}\,.
\end{align}
[To see this, choose a map $\lcssample_{\mathrm{I}}$ of rank $N_\gamma$ such that $\delta_\gamma(\Xi-\lcssample_{\mathrm{I}})=0$,
and set $\lcssample_{\mathrm{II}} := \lcssample- \lcssample_{\mathrm{I}}$.
It follows that $\delta_\gamma(\lcssample_{\mathrm{II}}) = 0$, so 
$r^{-\gamma} \|\sigma\|_r \to 0$ for all $ \sigma \in \img \lcssample_{\mathrm{II}}$.
Since Lemma~\ref{sigma_not_vanish_lemma} applies to $\lcssample_{\mathrm{I}}$,
the intersection of $\img \lcssample_{\mathrm{I}}$ with $\img \lcssample_{\mathrm{II}}$
is trivial; thus $\lcssample = \lcssample_{\mathrm{I}}  + \lcssample_{\mathrm{II}}$
is a direct sum.]
As a consequence hereof, and since $\delta_\gamma(\cdotarg) \leq \delta_{\gamma'} (\cdotarg)$
for $\gamma \leq \gamma'$, we obviously have $\Phi_\gamma \subset \Phi_{\gamma'}$ --
the $\Phi_\gamma$ form an increasing sequence of vector spaces.

We will show now that the elements of $\Phi_\gamma$ are Wightman fields.
To this end, we will approximate $\phi \in \Phi_\gamma$ with a sequence of
bounded operators localized in smaller and smaller regions. Our aim
is the following proposition.

\begin{prop} \label{convergenceprop}
Let $\phi \in \Phi_\gamma$. One can find a sequence of operators
$A_r \in \afk(r)$ ($r>0$) such that
\begin{equation*}
  \lnorm{ \phi - A_r }{\ell} \to 0 \quad
  \text{for some $\ell>0$}
\end{equation*}
{\em(}more precisely $\lnorm{ \phi - \psmap^\ast A_r }{\ell} \to 0${\em)}.
\end{prop}

\noindent
We will carry out -- or rather sketch -- the proof in four steps.
We choose $\lcssample \in \lcs$ of rank $N_\gamma$ such that
$\delta_\gamma(\psmap-\lcs)=0$ and set $S := \img \lcssample$.

\medskip
(a) First, consider the case of a rank-1 operator $\lcssample = \sigma \phi$.
According to Lemma~\ref{sigma_not_vanish_lemma}, we may choose a null sequence $\rho$
such that $\|\sigma\|_r \geq c \cdot r ^\gamma$ for some $c>0$, where $r \in \rho$.
For $r \in \rho$, we define $A_r$
as a linear form on the one-dimensional subspace 
of $\afk(r)_\ast$ spanned by $\sigma \restrict \afk(r)$,
namely by setting $\sigma(A_r) = 1$. We can then apply the Hahn-Banach theorem to
continue this form to an operator $A_r \in \afk(r) = (\afk(r)_\ast) ^\ast$; its norm
is bounded by
$\|A_r\| \leq (\|\sigma\|_r )^{-1}  \leq c^{-1} r^{-\gamma}$.
We must show that $\lnorm{ \phi - \psmap^\ast A_r }{\ell} \to 0$: This follows from
\begin{equation}
\lnorm{ \phi - \psmap^\ast A_r }{\ell}
= \lnorm{ \lcssample^\ast A_r - \psmap^\ast A_r }{\ell}
\leq \lnorm{ \lcssample - \psmap }{\ell} _r \, \|A_r\|
\leq {c}^{-1} r^{-\gamma} \, \lnorm{ \lcssample - \psmap }{\ell} _r \to 0
\end{equation}
for large $\ell$ and for $r \in \rho$. Thus $A_r \to \phi$ on $\rho$, and we may
easily find suitable $A_r$ for any $r > 0$ by choosing $A_r$ to be
``piecewise constant.''

\medskip
(b) Now let $\dim S > 1$, but let all $\sigma \in S$ have ``the same short distance
behavior,'' i.e., we assume that a function $\eta: \rbb^+ \to \rbb^+$ exists
such that
\begin{equation} \label{homogenous}
   c_1(\sigma) \eta(r) \geq \|\sigma \|_r \geq c_2(\sigma) \eta(r)
   \quad
   \text{for all $r \in \rho$, $\sigma \in S$},
\end{equation}
where $\rho$ is some fixed null sequence, and $c_1, c_2$ are positive constants depending
on $\sigma$. We also assume $\eta(r) \geq r^\gamma$ for $r \in \rho$.
It is easily seen that $c_1$, $c_2$ can be chosen to be locally uniform in $S$,
and thus we can choose them uniformly on the unit sphere of $S$ (with respect
to some fixed norm $\|\cdotarg\|_S$, independent of $r$): We can find constants $c_1', c_2'>0$ such that
\begin{equation} \label{homogenous2}
   c_1' \eta(r) \|\sigma \|_S  \geq \|\sigma \|_r \geq c_2' \eta(r) \|\sigma \|_S 
   \quad
   \text{for all $r \in \rho$, $\sigma \in S$}.
\end{equation}
Now let $\phi \in \Phi_\gamma$,
and write $\lcssample =  \sigma_0 \phi \oplus \lcssample'$ with some fixed $\sigma_0 \in S$
and $\rank \lcssample' = N_\gamma - 1$.
Again, we define $A_r$ $(r \in \rho)$ as a linear form on $S \restrict \afk(r)$
by
\begin{equation}
  \sigma_0(A_r) = 1, \quad
  A_r \restrict \img \lcssample' = 0,
\end{equation}
thus achieving $\lcssample^\ast A_r = \phi$. The norm of $A_r$ can be estimated as
\begin{equation}
  \| A_r \restrict S \| = \sup_{\sigma \in S} \frac{| \sigma(A_r)| }{ \|\sigma\|_r}
  \leq \frac{ \|A_r\|_S }{ c_2' \eta(r) } \leq \frac{c_3}{\eta(r)}
  \quad (r \in \rho)
\end{equation}
with some constant $c_3 > 0$. We may now extend $A_r$ to an element of
$\afk(r)$ and establish convergence using the methods outlined in (a).

\medskip

(c) More generally, let us assume that $S$ has the structure
\begin{equation} \label{ssum}
S = S_1 \oplus \ldots \oplus S_k
\end{equation}
with finite-dimensional spaces $S_j$ which fulfill conditions of the type \eqref{homogenous}
with respect to functions $\eta_j(r)$ and a common null sequence $\rho$;
moreover, we require $\eta_{j+1}(r) / \eta_j(r) \to 0$ and $\eta_k(r) \geq r^\gamma$ 
on $\rho$.
[For example, this is the situation met in free field theory, where the
functions $\eta_j(r)$ are powers of $r$, and $\rho$ is arbitrary.]
We write
$\lcssample = \lcssample_1 \oplus \ldots \oplus \lcssample_k$
with respect to the direct sum~\eqref{ssum},
and prove the proposition for $\phi \in \img \lcssample_j^\ast$ by induction on $j$.
Let the statement be true for all $j'<j$ in place of $j$ (we include the case $j=1$ here).
For $\phi \in \img \lcssample_j^\ast$,
define $A_r$ ($r \in \rho$) as a linear form on $S_j$
by $\lcssample_j^\ast A_r = \phi$; as in (b),
we have $\|A_r \restrict S_j\| \leq c / \eta_j(r)$ with some constant $c>0$,
and thus we can extend $A_r$ to an element of $\afk(r)$ with the same bounds on its norm.
Now observe that for $\ell$ sufficiently large,
\begin{multline}
  \lnorm{ \phi - \psmap^\ast A_r + \sum_{m=1}^{j-1} \lcssample_m^\ast A_r }{\ell}
 \leq \lnorm{ \lcssample^\ast A_r - \psmap^\ast A_r }{\ell}
  + \sum_{m=j+1}^{k} \lnorm{ \lcssample_m^\ast A_r  }{\ell}
  \\
 \leq  \lnorm{ \lcssample - \psmap }{\ell} _r \|A_r\|
  + \sum_{m=j+1} ^{k} \lnorm{ \lcssample_m }{\ell} _r  \|A_r\|
 \;\leq\; \lnorm{ \lcssample - \psmap }{\ell} _r \frac{c}{\eta_j(r)}
  + \sum_{m=j+1}^{k} \frac{\eta_m(r)}{\eta_j(r)} \cdot const \to 0 \quad (r \in \rho).
\end{multline}
Thus we conclude by induction,
\begin{equation}
\phi \in  \;
\overline{ \afk(r) + \bigoplus_{m=1}^{j-1} \img \lcssample_m^\ast }
\subset
\overline{ \afk(r) + \overline {\afk(r)}  } = \overline{ \afk(r) }
\quad \text{for any }r>0,
\end{equation}
where the bar denotes closure with respect to $\lnorm{\cdotarg}{\ell} $.

\medskip

(d) In the general case, the short distance behavior within $S$
may be more complicated than assumed above, and possibly depend crucially
on the choice of $\rho$.
Let $\rho$ be any fixed null sequence. Instead of considering
$\delta_\gamma(\cdotarg)$, we introduce a weaker
pseudometric $\delta_\gamma^\rho (\cdotarg)$
defined as in Eq.~\eqref{deltadef}, but restricting convergence to $r \in \rho$.
The results established so far for $\delta_\gamma(\cdotarg)$ and $\Phi_\gamma$
hold in an analogous way for $\delta_\gamma^\rho(\cdotarg)$ and
corresponding spaces $\Phi_\gamma^\rho \subset \Phi_\gamma$.
If $\rho' \subset \rho$ is a subsequence, then
$\Phi_\gamma^{\rho'} \subset \Phi_\gamma^\rho$; in case equality holds here
for all subsequences $\rho'$, we will call $\rho$ a {\em stable} sequence.
Let $\rho$ fulfill this condition, and choose $\rank \lcssample'$ minimal such that
$\delta_\gamma^\rho(\psmap-\lcssample') = 0$. By passing to subsequences 
of $\rho$, we can step by step enforce an ``ordering'' in 
the short distance behavior of the elements of $\img \lcssample'$,
and arrive at a situation as outlined in (c). Since $\rho$ is stable,
passing to subsequences does not affect the approximation property of $\lcssample'$;
so we can use the methods developed in (c) to
establish the proposition for $\phi \in \img \lcssample' \!\,^\ast =\Phi_\gamma^\rho$.

It remains to verify that $\cup_\rho \Phi_\gamma^\rho$ spans all of $\Phi_\gamma$, where
the union runs over all stable sequences $\rho$. To accomplish this,
assume that $\Phi_{\mathrm{st}} := \lspan (\cup_\rho \Phi_\gamma^\rho )$ is a
proper subspace of $\Phi_\gamma$. Choose $\phi \in \Phi_\gamma \backslash \Phi_{\mathrm{st}}$
and appropriate $\sigma_0 \in \Sigma$, $\lcssample_{\mathrm{st}} \in \lcs$ such that
\begin{equation} \label{psiSsplit}
  \lcssample = \lcssample_{\mathrm{st}} \oplus \sigma_0 \phi, \quad
   \img \lcssample_{\mathrm{st}}^\ast \supset \Phi_{\mathrm{st}}.
\end{equation}
According to Lemma~\ref{sigma_not_vanish_lemma}, we may choose $\rho$ such that
$\|\sigma_0\|_r \geq r^\gamma \cdot const$ for $r \in \rho$, where we may
take $\rho$ to be stable, possibly replacing it by a subsequence.
In analogy with \eqref{psisplit}, we split up $\lcssample$ as
\begin{align} \label{psiRhosplit}
 &\lcssample = \lcssample_{\mathrm{\mathrm{I}}} \oplus \lcssample_{\mathrm{II}}, \quad \text{where} \notag
 \\
 &\img \lcssample_{\mathrm{I}}^\ast = \Phi_\gamma^\rho, \;\quad
 r^{-\gamma} \|\sigma\|_r \xrightarrow[\rho]{\;} 0 \; \;\forall\, \sigma \in \img\lcssample_{\mathrm{II}}\,.
\end{align}
But since $\Phi_\gamma^\rho \subset \Phi_{\mathrm{st}}$, it follows
from Eqs.~\eqref{psiSsplit} and \eqref{psiRhosplit}
that $\sigma_0 \in \img \lcssample_{\mathrm{II}}$;
thus $r^{-\gamma} \|\sigma_0\|_r \to 0$ on $\rho$, which gives us a contradiction.
This completes the proof. \qed

\medskip
Having verified Proposition~\ref{convergenceprop}, we may now apply the results
of Fredenhagen and Hertel, \cite{FreHer:pointlike_fields} who considered
(when expressed in our notation) the following set of linear forms:
\begin{equation}
 \PhiFH := \big\{
   \phi \in \cinftys ^\ast \; \big| \;
   R^{\ell} \phi R^{\ell} \in \bigcap_{r>0} \overline{ R^{\ell} \afk(r) R^{\ell} }^w
   \text{ for some $\ell>0$}
 \big\}  ,
\end{equation}
where $\overline{\quad}^w$ denotes the weak closure.
$\PhiFH$ can be interpreted as the theory's field content.
The authors showed
that any $\phi \in \PhiFH$ is a local field associated with the net $\afk$;
more precisely,
\begin{equation}
  f \mapsto \phi(f) = \int f(x) \, U(x) \phi U(x)^\ast \, d^{s+1}x
\end{equation}
can be extended to an operator valued tempered distribution on the
domain $\cinfty = \bigcap_{\ell>0} R^\ell \hcal$, which is local and relatively local;
$\phi(f)$ is closable,
and its closure $\phi(f)^-$ is affiliated with the local algebras,
\begin{equation}
  \phi(f)^- \, \eta \, \afk(\ocal) \quad
  \text{if } \, \supp f \subset \ocal.
\end{equation}
In particular, these statements apply to the fields constructed here,
since Proposition~\ref{convergenceprop} shows that $\Phi_\gamma \subset \PhiFH$
for any $\gamma \geq 0$.
In fact, we will show that our spaces $\Phi_\gamma$ exhaust $\PhiFH$.
To this end, we first derive improved approximation properties
for the fields $\phi \in \PhiFH$.

\begin{lemm} \label{betterconvergencelemma}
Let $\phi \in \PhiFH$. One can find constants $\ell>0$, $k>0$ 
and operators $A_r \in \afk(r)$ for each $r>0$ such that 
\begin{equation*}
  \lnorm{A_r - \phi}{\ell} = O(r),
  \qquad
  \|A_r\| = O(r^{-k}).
\end{equation*}

\begin{proof}
We choose a test function $f \in \scal(\mkraum)$ with $f \geq 0$,
$\int f(x) d^{s+1}x = 1$, $\supp f \subset \ocal_{r=1}$,
and set $f_r := r^{-(s+1)} f(r^{-1}x)$.
Now let $r$ be fixed, and let $\phi(f_r)^- = V_r D_r$ be the polar
decomposition of this operator, where $V_r$ is a partial isometry,
and $D_r$ is self-adjoint. Since $\phi(f_r)^- \,\eta\, \afk(r)$,
both $V_r$ and all bounded functions of $D_r$ belong to $\afk(r)$.\cite{BraRobPolar} 
Let $\ell$ be sufficiently large
such that $\lnorm{\phi}{\ell} < \infty$; for $\epsilon > 0$, set
\begin{equation}
  A_{r,\epsilon} := \epsilon^{-1} V_r \sin (\epsilon D_r) \in \afk(r).
\end{equation}
Using the inequality
\begin{equation}
  \big(  x - \epsilon^{-1} \sin \epsilon x \big)^2
  \leq \epsilon^2 \, x^4 \quad
  \forall \, x \geq 0, \, \epsilon > 0 ,
\end{equation}
we can establish the estimate
\begin{equation} \label{aphiest}
\|\big(A_{r,\epsilon} - \phi(f_r) \big)R^{4\ell} \|^2
\leq \|  \big( \epsilon^{-1} \sin(\epsilon D_r) - D_r \big) R^{4\ell}  \|^2
\leq \epsilon^2  \|  D_r^2 R^{4\ell}  \|^2
= \epsilon^2 \| \phi(f_r)^\ast \phi(f_r) R^{4\ell} \|^2 .
\end{equation}
By repeated use of the relation
\begin{equation}
 \lbrack R, \phi(f_r)  \rbrack
 = - i R \,  \phi( \partial_0 f_r ) \, R
\end{equation}
[compare Eq.~(2.4) in Ref.~\onlinecite{FreHer:pointlike_fields}] and of the bound
$\lnorm{ \phi(g) }{\ell} \leq \int |g(x)| d^{s+1}x \cdot const$,
we can establish
\begin{equation}
  \| \phi(f_r)^\ast \phi(f_r) R^{4\ell} \| \leq r^{-4\ell}  c
\end{equation}
with some constant $c>0$ independent of $r$. Applying Eq.~\eqref{aphiest}, that yields
\begin{equation} \label{A_minus_phif_vanish}
  \lnorm{ A_{r,\epsilon} - \phi(f_r) }{ 4\ell } \leq \epsilon r^{-4\ell} c.
\end{equation}
Using the integral representation of $\phi(f_r)$ and the spectral properties
of the translation operators, one also verifies that
\begin{equation} \label{phif_minus_phi_vanish}
  \lnorm{ \phi(f_r) - \phi }{ \ell+1 } = O(r).
\end{equation}
Choosing now $A_r := A_{r,\epsilon}$ with $\epsilon = r ^{4\ell+1}$, 
we can combine \eqref{A_minus_phif_vanish} and \eqref{phif_minus_phi_vanish} to show
\begin{equation}
\lnorm { \phi - A_{r} }{ 4\ell +1} = O(r),
\end{equation}
which proves the lemma, since the required bounds on $\|A_{r}\|$ are obvious.
\end{proof}

\end{lemm}

\noindent
{\em Remarks:} Certainly, $\ell$ and $k$ can be chosen uniformly on any finite-dimensional 
subspace of $\PhiFH$. Furthermore, by again smearing $A_{r,\epsilon}$
with $f_r$ and rescaling, we may assume that $A_r$ is of the form $\hat A_r(f_r)$.

We are now in the position to prove that the $\Phi_\gamma$ exhaust $\PhiFH$.
To this end, let $\phi \in \PhiFH$. Choose a sequence $A_r$ as in
Lemma~\ref{betterconvergencelemma}, and fix $\gamma$ such that
$\|A_r\| =O(r ^{-\gamma})$. We will show that
$\phi \in \Phi_\gamma$: Choosing $\lcssample \in \lcs$, $\ell > 0$ such that
$\img \lcssample^\ast = \Phi_\gamma$ and $r^{-\gamma} \lnorm{\psmap-\lcssample}{\ell} _r \to 0$,
we can achieve
\begin{equation}
   \lnorm{\psmap^\ast A_r-\lcssample^\ast A_r}{\ell}  \to 0.
\end{equation}
Since $\lnorm{ \psmap^\ast A_r - \phi}{\ell} \to 0$ by construction
(if $\ell$ is sufficiently large), this means
\begin{equation} \label{phiapprox}
  \lcssample^\ast A_r \xrightarrow[r \to 0]{}  \phi \qquad
  \text{with respect to} \; \lnorm{\cdotarg}{\ell} .
\end{equation}
Note that the left-hand side of Eq.~\eqref{phiapprox} does not leave the finite-dimensional space
$\Phi_\gamma$, which is closed; thus $\phi \in \Phi_\gamma$. We have shown the following.

\begin{thm} \label{exhaustprop}
If the net $\afk$ satisfies the microscopic phase space condition, then
\begin{equation*}
  \PhiFH = \bigcup_{\gamma \geq 0} \Phi_\gamma \,.
\end{equation*}
\end{thm}

According to the results in Ref.~\onlinecite{FreHer:pointlike_fields}, this means
that our construction describes all local
fields affiliated with the net and satisfying polynomial $H$-bounds.

\section{Symmetry aspects} \label{symmsec}

In this section, we will investigate the action of symmetry transformations
on the pointlike fields we have constructed, and eventually show that
these fields satisfy the Wightman axioms.

First, we will revisit the structure connected with the spaces $\Phi_\gamma$
of local fields.
Given $\gamma \geq 0$, we choose a map $\lcssample\in \lcs$ of rank $N_\gamma$ such that
$\delta_\gamma(\psmap-\lcssample) = 0$; then $\img \lcssample^\ast = \Phi_\gamma$. We can define
the finite-dimensional space
\begin{equation}
  \Sigma_\gamma := \cinftys / \ker \lcssample;
\end{equation}
by virtue of Lemma~\ref{img_charact_lemma}, this definition is independent of $\lcssample$.
In a natural way, we have $\Sigma_\gamma^\ast = \Phi_\gamma$, and denoting
the canonical projection onto $\Sigma_\gamma$ by $p_\gamma$, its dual map
$p_\gamma^\ast$ is just the inclusion $\Phi_\gamma \hookrightarrow \cinftys^\ast$.
In summary, we get the following diagram, where
dashed lines connect pairs of dual spaces:
%
%
%
\begin{equation}   \label{dualDiagram}
\begin{picture}(100,60)(0,0)
  \put(10,50){\makebox(0,0)[c]{$\cinftys$}}
  \put(80,50){\makebox(0,0)[c]{$\Sigma_\gamma$}}
  \put(10,10){\makebox(0,0)[c]{$\cinftys^\ast$}}
  \put(80,10){\makebox(0,0)[c]{$\Phi_\gamma$}}
  \put(50,3){\makebox(0,0)[c]{$p_\gamma^\ast$}}
  \put(50,57){\makebox(0,0)[c]{$p_\gamma$}}
  \dashline{3}(10,20)(10,40)
  \dashline{3}(80,20)(80,40)
  \put(30,50){\line(1,0){40}}
  \put(30,10){\line(1,0){40}}
  \put(70,55){\arc(0,-5){-50}}
  \put(70,45){\arc(0,5){50}}
  \put(70,12){\arc(0,-2){180}}
  \put(30,15){\arc(0,-5){50}}
  \put(30,5){\arc(0,5){-50}}
\end{picture}
\end{equation}
The elements of $\Sigma_\gamma$ correspond to ``germs of states'' as
described by Haag and Ojima.\cite{HaaOji:germs}

The phase space approximation ``$\psmap \approx \lcssample$'' can be introduced into this
scheme in a simple way: Let $p:\cinftys^\ast \to \cinftys^\ast$ be an arbitrary projection
onto $\Phi_\gamma$, continuous with respect to the weak topology on $\cinftys^\ast$. 
Then the predual map $p_\ast$ exists, and since $\Phi_\gamma$ is finite-dimensional, we have 
\begin{equation}
\lnorm{p_\ast}{\ell} = \sup_{\sigma \in \cinftys} \frac{ \lnorm{p_\ast \sigma}{\ell} }{ \lnorm{ \sigma}{\ell} } < \infty
\end{equation}
for large $\ell$. This implies
%
\begin{equation}
   \lnorm{\psmap p_\ast - \psmap }{\ell} _r
  \leq  \lnorm{ (\psmap-\lcssample) p_\ast }{\ell} _r + \lnorm{ \psmap - \lcssample p_\ast }{\ell} _r
  \leq \lnorm{ \psmap-\lcssample }{\ell} _r \lnorm{p_\ast}{\ell} + \lnorm{ \psmap - \lcssample }{\ell} _r
  = o(r^\gamma),
\end{equation}
%
so we can express the approximation by
$\delta_\gamma(\psmap p_\ast - \psmap ) = 0$, without referring to a specific map~$\lcssample$.

Now we will turn to the investigation of symmetry operations. The action of
such symmetries (e.g., Lorentz transformations) is usually given on $\boundedops$,
and by virtue of diagram~\eqref{dualDiagram} we will transfer this action to
the spaces $\Phi_\gamma$. First of all, let us define the class of ``admissible
transformations'' that we will consider.

\begin{defn} \label{micro_symm_def}
Let $U \in \boundedops$ be a unitary. The transformation
$\alpha = \mathrm{ad\,} U : \boundedops \to \boundedops$
is called a {\em microscopic symmetry}
if the following two conditions hold:
\begin{enumerate}

\renewcommand{\theenumi}{(\arabic{enumi})}
\renewcommand{\labelenumi}{\theenumi}

\item \label{alpha_regular_cond}
  For every $\ell > 0$, there is an $\ell' > 0$ such that $\|R^{-\ell} U R^{\ell'}\| < \infty$.
  \item \label{alpha_local_cond}
  There are constants $c,R > 0$ such that
  $\alpha \afk(r) \subset \afk(c \cdot r) \; \forall r \in (0,R)$.
\end{enumerate}
\end{defn}
Condition \ref{alpha_regular_cond} allows us to extend $\alpha$ to a map
$\alpha: \cinftys^\ast \to \cinftys^\ast$. Regarding this extension, we prove the following.
\begin{prop}
  Let $\alpha$ be a microscopic symmetry. Then
  \begin{equation*}
    \alpha \Phi_\gamma \subset \Phi_\gamma \quad \forall \gamma \geq 0.
  \end{equation*}
\end{prop}

\begin{proof}
Using condition~\ref{alpha_regular_cond} in Definition~\ref{micro_symm_def},
we can define the predual map $\alpha_\ast: \cinftys \to \cinftys$. Let $\gamma \geq 0$
be fixed. Lemma~\ref{img_charact_lemma} shows us that in diagram~\eqref{dualDiagram},
\begin{equation}
  \ker p_\gamma = \big\{  \sigma \in \cinftys \, \big| \, r^{-\gamma} \|\sigma\|_r \to 0 \big\}.
\end{equation}
Together with condition~\ref{alpha_local_cond} in  Definition~\ref{micro_symm_def},
this yields $\alpha_\ast \ker p_\gamma \subset \ker p_\gamma$; thus $\alpha$ has a well-defined
action on the quotient space $\Sigma_\gamma$ and on its dual $\Phi_\gamma$.
By construction, this action is compatible with the inclusion
$p_\gamma^\ast: \Phi_\gamma \hookrightarrow \cinftys^\ast$,
which proves the proposition.
\end{proof}

Microscopic symmetries $\alpha$ thus
leave the spaces $\Phi_\gamma$ of pointlike fields stable.
In the case of a group representation $g \mapsto \alpha(g)$,
we get corresponding finite-dimensional representations
on every field space $\Phi_\gamma$. An example for this case are
Lorentz transformations: Each $\Phi_\gamma$ carries a finite-dimensional
representation of the Lorentz group. Another example for microscopic
symmetries are dilations, provided they exist as a symmetry of the net.
Furthermore, our analysis allows us to handle inner symmetries,
both broken and unbroken. In the case of a spontaneously broken
symmetry, one expects that $\alpha = \mathrm{ad\,} U$ preserves localization
only in regions of some limited size -- this corresponds to the case $R < \infty$ in
Definition~\ref{micro_symm_def}. [Compare Ref.~\onlinecite{BDLR:goldstone}.
To include the situation considered there in our context, it is necessary to
extend our framework from the observable algebras $\afk(\ocal)$ to the
field algebras $\ffk(\ocal)$, including also nonobservable fields,
which should however be straightforward.]

In extension of our methods introduced above, one observes that the antilinear
involution $A \mapsto A^\ast$ can be treated in a similar manner, showing that
the field spaces $\Phi_\gamma$ are invariant under Hermitean conjugation
$\phi \mapsto \phi^\ast$. Defining the translated fields
as $\phi(x) := U(x) \phi U(x)^\ast$, or passing over to the
``smeared'' fields $\phi(f)$,
we may also derive symmetry properties under the 
(connected part of the) full Poincaré group.

The only remaining part in establishing the Wightman axioms then
is irreducibility, i.e., the question whether the vacuum vector $\Omega$ 
is cyclic for the fields. As will be discussed below, we cannot expect this
in the general case. However, for each $\gamma$ we may consider 
a ``reduced'' Hilbert space,
\begin{equation}
    \hcal_\gamma := \overline{ \pcal_\gamma(\ocal) \Omega} \subset \hcal,
\end{equation}
where $\ocal$ is some open subset of $\rbb^{s+1}$,
and $\pcal_\gamma(\ocal)$ is the polynomial algebra generated by the
fields $\phi(f)$ with $\phi \in \Phi_\gamma$ and $\supp f \subset \ocal$.
(Due to the Reeh-Schlieder theorem, the space $\hcal_\gamma$ does not 
depend on $\ocal$.)
Considered on $\hcal_\gamma$, the set $\Phi_\gamma$ is certainly
irreducible. Thus, we have established the following.

\begin{thm}
Let $\ocal \mapsto \afk(\ocal)$ satisfy the microscopic phase space condition.
For every $\gamma \geq 0$, there exists
a basis $\{ \phi_1, \ldots, \phi_{N_\gamma} \}$ of $\Phi_\gamma$ and a Hilbert space
$\hcal_\gamma \subset \hcal$ such that
$\{ f \mapsto \phi_j(f) \}$ is a set of quantum fields on $\hcal_\gamma$ in the sense of
the Wightman axioms.
\end{thm}

Here we refer to the Wightman axioms as put forward in Ref.~\onlinecite{StrWig:PCT}.
Note that in our case, the fields always have finite spin. In the more general
framework of Ref.~\onlinecite{FreHer:pointlike_fields}, this is not guaranteed.
Also note that we need to allow $\hcal_\gamma$ to grow with $\gamma$: While at any
fixed $\gamma$, we are sure to find only a finite number of fields, more and more
independent fields may occur in the phase space approximation
as we increase the energy dimension.

Let us briefly return to the question whether, or on which Hilbert space,
the entire field content $\PhiFH$ is irreducible. In general, we cannot expect $\PhiFH$
to be an irreducible set on $\hcal$, since there exist nontrivial theories
which fulfill the microscopic phase space condition, but have a 
trivial field content, i.e., $\PhiFH = \cbb \idop$ (cf. the Appendix). 
These theories might always occur as a tensor factor of $\afk$.
To exclude such nonpointlike components from the theory, we can define the following subnet
$\afk_F$ of $\afk$ using the methods exposed in Ref.~\onlinecite{DSW:fields_algebras}:
\begin{equation}
  \afk_F (\ocal) :=   \pcal(\ocal) '',
\end{equation}
where $\pcal(\ocal)$ is the polynomial algebra generated by $\PhiFH$,
and $(\ldots)'$ denotes the weak commutant.
If $\afk$ fulfills the microscopic phase space condition, then so does $\afk_F$,
leading to the same field content as $\afk$; thus we may call $\afk_F$
the point field part of the theory. It is 
$\hcal_F :=  \overline{ \pcal(\ocal) \Omega }$ which we should consider
as the natural Hilbert space for the Wightman fields. 
In fact, using the techniques of Bisognano
and Wichmann,\cite{BisWic} 
it should be possible to show that
\begin{equation}
   \hcal_F = \hcal  \equivalent  \afk_F  = \afk,
\end{equation}
meaning that $\PhiFH$ is irreducible on $\hcal$ if, and only if, 
the theory is completely determined by pointlike observables.

We can incorporate derivatives of fields into our context as well.
It is easy to see that together with $\phi$, its derivatives
\begin{equation}
\partial_\mu \phi :=
\frac{\partial}{\partial x^\mu} U(x) \phi U(x)^\ast \big|_{x=0} = i [P_\mu,\phi]
\end{equation}
are contained in
$\PhiFH$; so we can take $\partial_\mu$ to be a linear operator from $\Phi_\gamma$
into some $\Phi_{\gamma'}$.
Usually, we must choose $\gamma' > \gamma$ here, since the ``energy dimension''
of fields increases when applying time derivatives:
If $\lnorm{\partial_0^n \phi}{\ell} < \infty $ for some fixed~$\ell$ and {\em any} $n \in \nbb$,
then it follows that $\| \phi \Omega \| < \infty$; however, this is only
possible if $\phi$ is a multiple of the identity. \cite{BucVerNobound}
In particular, the case of a finite-dimensional $\PhiFH$
automatically leads to $\PhiFH = \cbb \idop$.

\section{Conclusions and Outlook} \label{outlooksec}

Starting from a relativistic quantum theory in the algebraic framework, i.e.,
expressed in terms of observables localized at finite distances, we have shown
that its field content can be characterized by its phase space structure
in the short distance limit. We have introduced a 
physically motivated phase space criterion that
distinguishes a class of models with ``regular'' short distance behavior. 
Assuming that this
criterion is fulfilled, we have established very detailed results on the approximation
of pointlike fields by bounded local observables. The field content is exhausted
by an increasing sequence of finite-dimensional spaces $\Phi_\gamma$, each
of which is invariant under Lorentz transformations, Hermitean conjugation and
other symmetries. Their dimension $N_\gamma = \dim \Phi_\gamma$
can be read off directly from the phase space behavior.
The label $\gamma$ may be interpreted as a ``short distance dimension''
of the fields.

In mathematical terms, we have developed a method of classifying the short
distance behavior of a net of algebras $\ocal \mapsto \afk( \ocal)$;
the field space dimensions $N_\gamma$
are ``invariants'' of the net. Such a classification should depend on local properties
at small distances only. Strictly speaking, this goal is not completely reached
in our analysis: We incorporate the energy operator
$H$ as a global property. However, $H$ enters the construction not as the
generator of a global symmetry, but only through the more qualitative feature
of energy damping, and via its rôle as a generator of the unitary  group
used for ``smearing'' the fields.
One should therefore be able to replace the Hamiltonian with the local symmetry
generators established by Buchholz, D'Antoni, and Longo \cite{BDL:noether_thm}
through the use of a ``universal localizing map.''
It is also worth noting that with respect to certain phase space
properties, $H$ may be replaced\cite{BDL:modular_nuclearity}
with a suitable modular
operator $\Delta$. This might point to
an extension of our results even to situations without any space-time symmetries.
Since $- \log \Delta$ corresponds to energy-momentum {\em transfer} rather than
to total energy, though, it is unclear how the concept of polynomial bounds
can be substituted.


Besides a characterization of the short distance limit, our results provide
a sufficient technical basis for a rigorous proof of the existence of
operator product expansions; this will be discussed
in a forthcoming paper.\cite{Bos:products-article}

\begin{acknowledgments}
The author wishes to thank D. Buchholz for posing the problem,
and is grateful to him for many helpful discussions and remarks.
The author would also like to thank R.~Haag and I.~Ojima for a discussion 
on the subject, as well as H.~Reeh for his support in the beginning of the 
work that lead to the present paper.
The work has profited from
financial support by Evangelisches Studienwerk, Villigst,
which the author gratefully acknowledges.
\end{acknowledgments}


\appendix*

\section{Some simple models}

\renewcommand{\thedefn}{\Alph{section}.\arabic{defn}}
\renewcommand{\theequation}{\Alph{section}\arabic{equation}}

The claim that our phase space condition should hold in physically relevant theories
may be supported by the fact that it is fulfilled at least in some simple situations.
In this appendix, we will therefore investigate 
the structures introduced in the main text
in specific noninteracting models.

We will first consider a real scalar free field
and argue that this theory fulfills the microscopic phase space condition
in $s \geq 3$ spatial dimensions.
Then we discuss extensions of this result to more general
(still noninteracting) situations.

\subsection*{The real scalar free field}   

Most of the relevant techniques that allow us to discuss the 
short distance structure of free fields 
are already visible in the case of a single real scalar
field. For this theory, the following theorem holds.

\begin{thm} \label{rsffThm}
The theory of a real scalar free field of mass $m \geq 0$ in $s+1$ space-time
dimensions, $s \geq 3$, satisfies the microscopic phase space condition.
\end{thm}

This result was proved in Chap.~7 of Ref.~\onlinecite{Bos:Operatorprodukte} for a slightly
modified version of the phase space condition. 
The proof carries over to our situation; however, the construction
is quite involved, and it would go beyond the scope of the current paper 
to reproduce the complete discussion. Instead, we shall confine ourselves
to a rough and somewhat heuristic sketch of the arguments; 
the reader is referred to Ref.~\onlinecite{Bos:Operatorprodukte} for details.

In order to fix our notation, we shall briefly recall the definition of a real
scalar free field. Our Hilbert space $\hcal$ is the symmetric
Fock space over the single particle space $\kcal = \lcal^2 (\rbb^s,d^s p )$
(``wave functions in momentum space''). The scalar product on $\kcal$ will be
denoted as $ \etskp{ \cdotarg }{\cdotarg } $, as opposed 
to the scalar product $\hrskp{\cdotarg}{\cdotarg}$ on $\hcal$.
On $\kcal$, we have the generator
of time translations $\omega = \sqrt{ \pvq + m^2 }$, which fixes the Hamiltonian
$H$ on Fock space by means of ``second quantization.'' 
The spectral projectors of $\omega$ will be denoted as $P_\omega(E)$, those of
$H$ as $P_H(E)$.

The local algebras $\afk(r)$ are then generated by Weyl operators 
$W(f) = \exp i (a(f) + a\st(f))$ with $f=f^+ + i f^-$,
where $f^\pm \in \kcal$ are localized functions in configuration space,
in the sense that the Fourier transforms of
$\opmh f^\pm$ are real-valued and smooth and have their support within the ball $|\xv| < r$.
 
In order to prove Theorem~\ref{rsffThm}, we must find a series
expansion of the kind
\begin{equation} \label{XiExpHeuristic}
  \Xi = \sum_j \sigma_j \phi_j 
\end{equation}
with linear forms $\phi_j \in \cinftys^\ast$ and functionals $\sigma_j \in \Sigma$,
valid for localized arguments of $\sigma_j$ and energy-damped expectation values
of $\phi_j$. We will establish this expansion in four steps.

\paragraph*{Step 1: Single particle space.}

\newcommand{\chiE}{\hat \chi_E}
\newcommand{\chiR}{\chi_r}

For a start, we will derive a series
expansion of the single-particle scalar product
\begin{equation} \label{productForExpansion}
  \etskp{ \fpm }{ k } = \etskp{ \opmh \fpm }{ \omph k },
\end{equation}
where $f^\pm$ is localized with radius $r$ (as above), and $k$ is an energy-bounded
function, i.e. $k \in P_\omega(E) \kcal$ for some $E$. 
Since $\omph k$ is also energy bounded, its Fourier transform 
$\fcal \lbrack \omph  k \rbrack$ is analytic, 
and thus we can replace it with its Taylor series,
\begin{equation}
   \fcal \big\lbrack \omph  k \big \rbrack (\xv) =
   \sum_{ \kappa } \frac{1}{\kappa!}
                    \frac{\partial^\kappa}{\partial x^\kappa}
                  \fcal \big\lbrack \omph  k \big \rbrack \Big|_{\xv=0}
                  x^\kappa
\end{equation}
-- the sum runs over multi-indices $\kappa$ here. Inserting
this expansion into Eq.~\eqref{productForExpansion}, and rewriting the spatial derivatives
as integrals in momentum space, we arrive at
\begin{equation}\label{etskpSeries1}
\etskp{\fpm}{k} = \sum_\kappa \etskp{\fpm}{
          \underbrace{ \frac{\sqrt{2}}{\kappa!}\, \opmh \fcal^{-1}[x^\kappa]}
                _{=: h^\pm_\kappa}  }
           \etskp{ \underbrace{
 \frac{i^{-|\kappa|}}{\sqrt{2} \, (2\pi)^{s/2}} \omph p^\kappa}
         _{=: g^\pm_\kappa}}{ k }
%
        = \sum_\kappa \etskp{f^\pm}{h_\kappa^\pm} \etskp{g_\kappa^\pm}{k}
.
\end{equation}
The scalar products with the ``improper vectors'' 
$g_\kappa^\pm$ and $h_\kappa^\pm$ can be justified due to the given
localization properties of $\fpm$ and $k$.
Note that $g_\kappa^\pm$ and $h_\kappa^\pm$ are independent  
not only of $\fpm$ and $k$, but also of $E$ and $r$. 
Using that the coefficients $\etskp{f^\pm}{h_\kappa^\pm}$ are real, we can thus write
\begin{equation}\label{etskpSeries2}
\fpm = \sum_\kappa \etskp{f^\pm}{h_\kappa^\pm} g_\kappa,
\end{equation}
where the sum is to be read ``under energy restriction.''

We note that regarding the short-distance or high-energy behavior
of the functions introduced above, one has
\begin{equation} \label{stEstimates}
   \| P_\omega(E) g_\kappa^\pm \| \sim E^{|\kappa| + (s \mp 1)/2},
   \quad
   | \etskp{\fpm}{h_\kappa^\pm} | \sim r^ {|\kappa| + (s \mp 1)/2} \; ;
\end{equation}
in particular, there are only finitely many terms in the sum 
\eqref{etskpSeries1} which correspond to a given ``scaling dimension.''

\paragraph*{Step 2: Expansion of Weyl operators.} 

In order to transfer our results in single particle space to the theory in
Fock space, we will next aim at a series expansion of 
local Weyl operators $W(f)$, where $f=f^+ + i f^-$ as before.
We can certainly expand them as
\begin{equation} \label{WeylESumme}
W(f) = e^{-\|f\|^2 /2} e^{i a\st(f)} e^{i a(f)}
%
     = e^{-\|f\|^2/2} 
     \!\!\!\!\!
     \sum_{m^\pm, n^\pm \in \nbb \cup \{0\}}
          \frac{i^{m^+ + n^+ + 2 m^- }}{m^+! m^-! n^+! n^-!}
          \, a\st(f^+)^{m^+} a\st(f^-)^{m^-} a(f^+)^{n^+} a(f^-)^{n^-}.
\end{equation}
If this sum is evaluated in energy-restricted states, we can 
insert our result \eqref{etskpSeries2} here, writing, for example,
\begin{equation}
   a\st(f^+) = \sum_\kappa \etskp{f^+}{h_\kappa^+} \, a\st(g_\kappa^+).
\end{equation}
However, this leaves us with additional $m^+ + m^- + n^+ + n^-$  
summation (multi-)indices in each summand of \eqref{WeylESumme}. 
In order to simplify the notation, we will reorganize this multiple sum by the following:
\begin{enumerate}

\renewcommand{\theenumi}{(\roman{enumi})}
\renewcommand{\labelenumi}{\theenumi}

\item relabeling the functions $g_\kappa^\pm$ and $h_\kappa^\pm$ with natural numbers $j$ instead of multi-indices $\kappa$,
\item grouping terms with equal powers of $ \etskp{f^+}{h_j^+} $
    and $ \etskp{f^-}{h_j^-} $ into one, and
\item labeling these terms with two multi-indices $\mu^\pm = (\mu_1^\pm,\ldots)$,
where $\mu_1^+$ counts the power of $ \etskp{f^+}{h_1^+} $, etc.
Additionally, we will combine $\mu^+$ and $\mu^-$ into a single multi-index $\mu = (\mu^+,\mu^-)$.
\end{enumerate}
This leads us to an expression
\begin{equation}\label{WfExpansion1}
 W(f) = \sum_{\mu} e^{-\|f\|^2/2}
          \prod_j \etskp{f^+}{h_j^+} ^{\mu_j^+} \etskp{f^-}{h_j^-} ^{\mu_j^-}
          \cdot \phi_{\mu},
\end{equation}
where $\phi_{\mu}$ are quadratic forms of the sort
\begin{equation} \label{phiDefSumme}
 \phi_{\mu} = \sum \frac{i^{m^++n^++2m^-}}{m^+!m^-!n^+!n^-!}
                     a\st(g_?) \ldots a(g_?) \ldots \; .
\end{equation}
Here $a\st(g_?) \ldots a(g_?) \ldots $ are certain products of annihilation
and creation operators of $g_j^\pm$, with their multiplicity given by $\mu^\pm$.
The $\phi_\mu$ can be shown to be elements of $\cinftys\st$. Their
detailed structure is not relevant for our purposes -- 
see, however, the end of this section for some examples.
Regarding the scaling properties of the expansion terms, we find
\begin{equation}
   \| P_H(E) \phi_\mu P_H(E) \| \sim E^{\vartheta(\mu)},
   \quad
   \prod_j \etskp{f^+}{h_j^+} ^{\mu_j^+} \etskp{f^-}{h_j^-} ^{\mu_j^-}
     \sim r^{\vartheta(\mu)}.
\end{equation}
The exponent $\vartheta(\mu)$ results from Eq.~\eqref{stEstimates},
and its detailed form is not of much importance; the
crucial point is that for every $\gamma \geq 0$, the set
$\{ \mu | \vartheta(\mu) \leq \gamma  \}$ is finite, given that $s \geq 2$.

\paragraph*{Step 3: Linear extension.} 

Our next step is to generalize the expansion \eqref{WfExpansion1}
to arbitrary $A \in \afk(r)$ in place of the generating $W(f)$. To that end,
we need to replace the numerical factors involving $\fpm$ with an expression
that is linear in $W(f)$; i.e., we need to find linear functionals $\sigma_\mu$
such that
\begin{equation} \label{sigmatarget}
\sigma_\mu \big( W(f) \big) =
e^{-\|f\|^2/2}
          \prod_j \etskp{f^+}{h_j^+} ^{\mu_j^+} \etskp{f^-}{h_j^-} ^{\mu_j^-}.
\end{equation}
In fact, we note that functionals of the type
\begin{equation}
  \sigma (\cdotarg) = \bighrskp{ \Omega }{ \; [a(b), \, [a^\ast (b'), \cdotarg ] \, ] \; \Omega }
  \qquad \text{with } b,b' \in \kcal
\end{equation}
evaluate on Weyl operators to
\begin{equation}
  \sigma \big( W(f) \big) =
  e^{- \|f\|^2/2}  \etskp{b}{i f} \, \etskp{ i f }{ b' } .
\end{equation}
For a given multi-index $\mu = (\mu^+,\mu^-)$, we can construct a functional
$\tau_\mu$ in a similar way such that for given $b_j,b_j' \in \kcal$,
\begin{equation} \label{taunudef}
  \tau_\mu \big( W(f) \big) =
  e^{- \|f\|^2/2}  \prod_j \etskp{b_j}{i f } ^{\mu_j^+}
   \etskp{ i f }{ b_j' } ^{\mu_j^-}.
\end{equation}
This is not exactly the form \eqref{sigmatarget} we need. However,
inserting $f = f^+ + i f^-$, 
we may build certain linear combinations of the $\tau_\mu$ in order
to form the $\sigma_\mu$ we require; this combinatorial problem
can be solved by means of a generating function technique,
which we shall not present in detail here.

After establishing the functionals $\sigma_\mu$, we can rewrite
\eqref{WfExpansion1} as
\begin{equation} \label{XiMuSum}
    \Xi   = \sum_{\mu}  \sigma_\mu \phi_\mu,
\end{equation}
where, at this time, the series is meaningful on a sufficiently restricted domain.

\paragraph*{Step 4: Convergence.}

In order to establish the phase space condition, our objective is to show that
the series expansion of $\Xi$ in Eq.~\eqref{XiMuSum}
holds with respect to the pseudometric $\hat \delta$ defined in Eq.~\eqref{hatdeltadef}. 
Loosely speaking, we have dealt with the algebraic aspects 
of this expansion up to now, while we still must handle the
topological problems, i.e., establish convergence. Roughly, this includes
the following tasks.

First, we need to establish precise estimates for the norms of $\phi_\mu$ and
$\sigma_\mu$, depending on the energy and configuration space localization
$E$ and $r$. This involves all parts of the calculation in steps 1--3 above.
The goal is to establish an estimate of the form
\begin{equation}\label{sigmaPhiEst}
	\|\sigma_\mu \restrict \afk(r) \| \cdot  \|  P_H(E)  \phi_\mu P_H(E) \|
	\leq c_\mu (Er)^{\vartheta(\mu)},
\end{equation}
where the constants $c_\mu$ do not increase too fast when varying $\mu$.

Second, we need to sum the terms in Eq.~\eqref{sigmaPhiEst} in order to establish norm 
convergence of the series \eqref{XiMuSum}  at fixed $E$ and $r$. Unfortunately, it turns out
that the estimates which we can establish in Eq.~\eqref{sigmaPhiEst} 
are not strict enough: 
We encounter convergence problems at ``high particle numbers,'' i.e.,
at high values of $|\mu|$. This problem is solved by handling the terms
for high $|\mu|$ according to a different expansion with better convergence
properties, using the techniques developed by Buchholz and Porrmann.\cite{BucPor:phase_space}
Their methods, however, result in expansion terms that are explicitly 
dependent of $E$ and $r$, so that they are useful for discussing
the convergence issues in question, but cannot be used in our main expansion.
In this part of the construction, the condition $s \geq 3$ is needed.

Third, once that the series is established at fixed $E$ and $r$, 
we need to apply the results
to the pseudometrics $\delta_\gamma$, which means in particular
to pass from the sharp energy bounds to the polynomial bounds used in the main text,
and to consider the limit $r \to 0$. As a result of this calculation,
we can show that for each $\gamma \geq 0$,
\begin{equation}
  \delta_\gamma \Big( \psmap
	- \sum_{ \vartheta(\mu) \leq \gamma }  \sigma_\mu \phi_\mu\Big) = 0.
\end{equation}
This relation finally proves Theorem~\ref{rsffThm}.

It is instructive to see how the approximation terms $\sigma_\mu \phi_\mu$ 
are formed explicitly. When working out the details of the construction,
one easily sees that the first term of the expansion,
corresponding to $\mu = (0,0,0,\ldots)$, leads to $\phi_\mu = \idop$ and
$\sigma_\mu = \hrskp{\Omega}{ \cdotarg | \Omega}$.
In the next term [with the next lowest value of $\vartheta(\mu)$] 
the form $\phi_\mu$ is the usual scalar field $\phi(0)$. 
Higher-order terms involve derivatives of the field and Wick products.
For $s=3$, the beginning of the expansion reads
\newcommand{\expansionline}[3]{
 \mbox{
  \parbox[c]{1.5cm}{
    \begin{flushright}
    \begin{math}
      \displaystyle #1
    \end{math}
    \end{flushright}
  }
  \parbox[c]{6cm}{
     \begin{math}
     	\displaystyle #2
     \end{math}
  }
  \parbox[c]{1.5cm} {
     \begin{math}
     	\displaystyle #3
     \end{math}
  }
 }
}
\begin{equation} \label{XiKonkretEntw2}
\begin{array}{l}
\left.
\expansionline
 {\psmap = \quad}
 {\hrskp{ \Omega }{ \,\cdot\, | \, \Omega } }
 {\cdot \; \idop }
\right\} \vartheta(\mu) = 0
\\
\left.
\expansionline
 {+}
 {\half \Big( \hrskp{h}{ \,\cdot\, |  \Omega } + \hrskp{\Omega}{ \,\cdot\, |  h} \Big) }
 {\cdot \; \phi(0) }
\right\} \vartheta(\mu) = 1
\\
\!\!\left.
\mbox{
 \begin{minipage}{9.38cm}
 \begin{flushleft}
 \expansionline
 {+}
 {\half \sum_{j=1}^3
   \Big( \hrskp{ \hat x_j h }{ \,\cdot\, |  \Omega }
  + \hrskp{\Omega}{ \cdotarg |  \hat  x_j h } \Big) }
 { \cdot \; \partial_j \phi(0) }
  \\
  \expansionline
 {+}
 {\frac{1}{2i} \Big( \hrskp{\omega^{-1} h }{ \,\cdot\, |  \Omega } -
       \hrskp{\Omega}{ \cdotarg |  \omega^{-1}  h} \Big) }
 { \cdot \; \partial_0 \phi(0) }
  \\
  \expansionline
 {+}
 {\sigma_Q(\cdotarg)}
 { \cdot\;  \mathrm{:}\phi^2\mathrm{:}(0) }
 \end{flushleft}
 \end{minipage}
}
\right\} \vartheta(\mu) = 2
\\
\left.
\expansionline
 {+}
 {\cdots.}
 {}
\right.
 \\
\end{array}
\end{equation}
Here $h=h_{\kappa=0}$ is the function introduced in \eqref{etskpSeries1},
the operator $\hat x_j$ stands for $\omega^{1/2} \fcal^{-1}x_j \fcal \omega^{-1/2}$,
with $\fcal$ being the operator of Fourier transformation,
and $\sigma_Q$ denotes the following functional:
\begin{equation}
\sigma_Q(\cdotarg) =
\frac{1}{4 \sqrt{2} }  \Big(\hrskp{h \otimes h}{ \cdotarg   |\Omega }
+ \hrskp{\Omega}{ \cdotarg  |h \otimes h } \Big)
+ \frac{1}{4 }  \hrskp{h}{  \cdotarg   |h }
- \frac{ \|h\|^2 }{4}  \hrskp{\Omega}{  \cdotarg   |\Omega } \,.
\end{equation}
We have $N_0 = 1$, $N_1=2$, $N_2=7$, and $N_3 = 21$ 
[the latter not being shown in Eq.~\eqref{XiKonkretEntw2}].
The corresponding spaces $\Phi_\gamma$ have the following form:
\begin{align}
\notag
\Phi_0 &=
	\lspan\{\idop\},
\\
\notag
\Phi_1 &=
	\lspan \{\idop, \, \phi \},
\\
\Phi_2 &=
	\lspan \{ \idop, \,  \phi, \, \partial_j \phi, \, \partial_0 \phi, \, \wickprod{\phi^2} \},
\\
\notag
\Phi_3 &=
	\lspan \{ \idop, \,  \phi, \, \partial_j \phi, \, \partial_0 \phi, \, \wickprod{\phi^2},
	      \, \partial_j^2 \phi, \, \partial_j \partial_k \phi, \, \partial_0 \partial_j \phi, 
	      \, \wickprod{\phi \;\partial_j \phi}, \, \wickprod{\phi \; \partial_0 \phi}, 
	      \, \wickprod{\phi^3} \},
\end{align}
where $j,k = 1 \ldots 3$, $j > k$, and the argument of the field has been omitted.
Note that the second-order time derivative $\partial_0^2 \phi$ lies in $\Phi_3$, 
but is a linear combination of the basis elements listed above 
-- this reflects the field equation of the model.

\subsection*{More applications}  

The methods we have sketched above are not restricted to the simple case
of a real scalar field.
Similar results can be derived in a large class of free theories, including
\begin{enumerate}

\renewcommand{\theenumi}{(\roman{enumi})}
\renewcommand{\labelenumi}{\theenumi}

\item
charged fields and fields of higher spin,

\item
theories with more than one field (but finitely many), massive or massless,
in $s \geq 3$ dimensions,

\item
massive theories in $s=2$,

\end{enumerate}
where on the technical side, only the single particle space expansions 
must be adapted appropriately.

It has not yet been possible to establish the criterion in the ($2+1$)-dimensional
massless case. This results from the model's peculiar infrared structure that causes
certain nuclearity conditions to be violated (cf.~Ref.~\onlinecite{BucPor:phase_space}),
leading to convergence problems at high particle numbers. Also, there is no result
for the ($1+1$)-dimensional case (irrespective of $m$); here the Wick powers
$ \wickprod{\phi^n} $ have the same energy dimension as the field itself, thus
one would not expect finite-dimensional field spaces $\Phi_\gamma$ to exist.

The criterion is explicitly violated in theories with infinitely many
free fields (see Sec.~8.1 of Ref.~\onlinecite{Bos:Operatorprodukte}).
In this case, we find $N_\gamma = \infty$ for all $\gamma \geq 1$.

In conclusion, we briefly sketch a model discussed by Lutz, \cite{Lut:UVFix}
which is interesting in this context since its field content is trivial.
We proceed from a free massless theory in
$(s+1)+1$ space-time dimensions, with the local algebras 
generated by Weyl operators $W(f)$ as before,
but with test functions $f$ of the type
\begin{equation}
  f (\pv) = (p_{s+1})^{2 n(r)} f_0(\pv),
\end{equation}
where $n(r)$ is an integer function tending to infinity as $r \to 0$, and $x_{s+1}$ is the
``additional'' spatial coordinate. When restricting the net to
$s+1$ space-time dimensions, we can apply the above arguments to show
that the theory fulfills the microscopic phase space condition. Indeed, we see
from \eqref{etskpSeries1} that every fixed expansion term $h_\kappa$ in the single particle space
simply drops out if $n(r)$ is sufficiently large, i.e., if $r$ is sufficiently small.
This leads to the result
\begin{equation}
  \delta_\gamma \big( \psmap - \hrskp{\Omega}{ \cdotarg | \Omega}  \, \idop  \big) = 0
  \quad
  \forall \gamma \geq 0,
\end{equation}
which implies $\Phi_\gamma = \cbb \idop$ for all $\gamma$.

\end{document}